\documentclass[aps,prd,superscriptaddress,eqsecnum,amsfonts,showpacs,epsfig]{article}
\usepackage{epsfig}

\newcommand{\be}{\begin{equation}}
\newcommand{\ee}{\end{equation}}
\newcommand{\bea}{\begin{eqnarray}}
\newcommand{\eea}{\end{eqnarray}}

\newcommand{\nn}{\nonumber}
\newcommand{\ep}{i\epsilon}

\begin{document}

%\preprint{ \parbox{1.5in}{\leftline{hep-th/000000}}}

\title{Excited  s-wave $1^{--}$ vector mesons, their leptonic decays and (in-)complete absence of abnormal states as seen from the constituent quark BSE}

\author{V. \v{S}auli
\\
DPT NPI Rez near Prague, Czech Academy of Sciences  }
\maketitle

\begin{abstract}
Within a lattice inspired  interaction between quark and antiquark, we obtain a hierarchy of solutions to the Bethe-Salpeter equation (BSE) 
for vector quarkonia excited states in the constituent quark mass approximation. 
As a toy model, we apply the similar  to calculate ground and excited states of $\phi$  and $\omega/\rho$ meson.
Through detailed numerical searches, our study provides evidence that the single-valued constant mass quark propagator does not yield known meson spectra without the simultaneous presence of abnormal (unphysical) solutions.
We classify normal and abnormal solutions and discuss necessary changes in the calculation scheme to avoid the spectrum of inconsistent solutions.     
While all experimental narrow vector mesons are identified with a normal state of the BSE, the occurrence of mutually cancelled
normal-abnormal states are reported. The results are slightly different for the heavy and light mesons, but in both cases they seem to result from the inconsistent use of the hmogeneous BSE to describe broad resonances.    
\end{abstract}

%\maketitle

\section{Introduction}
The concept of the constituent quark has its historical place in the development of our understanding of QCD. It is thought to be approximately valid for the description
of phenomena where the quantum effects responsible for the running quark masses can be ignored. 
Light quarks with a constituent mass $ m_u\simeq m_d \simeq 300 MeV$ can be used to estimate nucleon properties, such as mass and magnetic moments. 
This concept works roughly over the entire baryonic sector, while a similar attempt to describe light pseudoscalar mesons has failed, 
due to their Nambu-Goldstone boson character.  

 For heavier mesons, the constituent quark (CQ) approximation (CQA) is more appropriate
\cite{QR1977,EGKLY1980} with model-dependent constituent charm (bottom) quark mass in the range $1.3-1.8 (4.2-5.2) GeV$   
\cite{BPSV2005,SOEF2016,AN2020,SGPP2020,CR2022,BCVW2023,Br2023,AW2023,CD2024,K2024,OEFS2024,KL2024,Z2024}.

  The instantaneous Salpeter \cite{CH2018,GBN2019,WW2020,GGB2021} and quasipotential \cite{LLMPSVB2017,LSPB2017,FGK2020} approximations
   are the most widely used three-dimensional reductions of BSE.  Similar to methods based on light-cone Hamiltonians,
 they often exploit the CQ concept. 
Extended by the phenomenological confinement interaction, these approaches yield hadron masses and leptonic decays in good agreement with experiment. It is thought that the proximity of the sum of CQ masses to the physical mass of hadrons makes CQA particularly viable. 

Here we provide evidence that a closer look reveals a somewhat more complicated picture. Using a relativistic description based on quantum field theory
we will report on the inherent inconsistency of treating broad resonances in the naive way we treat excited states in quantum mechanics.

We will show that the homogeneous Bethe-Salpeter equation (BSE) provides a more complicated description when applied naively to the  
excited strangeonium and the complex of $\omega$ and $\rho$ excitations. While CQA BSE works reasonably well for describing heavy quarkonia, it fails when trying to describe objects we sometimes call excited light vector mesons, i.e. $\omega'$ $\omega^{''},. \rho^{'}...$, we get the abnormal solutions in addition to the normal states. Within the achieved numerical accuracy we observe that the poles of the abnormal states cancel against the normal solution.
In this paper we explain how the width of the assumed bound state or resonance is reflected or ignored by the BSE solutions. 
or ignored by the BSE solutions. Furthermore, we explain under which condition the solitary abnormal states disappear from the solution in CQA BSE.   

Before presenting the results, we give a brief overview of the achievements in the Dyson-Schwigner (DS) formalism.  
Within the DS/BS formalism, the constituent mass concept was challenged in \cite{S2010},
  where BSE with constituent quark propagators was compared with the more elaborate Maris-Tandy model (MTM) \cite{MATA1999}.
  In the latter model, the propagators were calculated from corresponding gap equations.
The MTM is the most studied phenomenological BSE model for mesons to date, it is based on the ladder rainbow truncation and was originally developed to fit 
light pseudoscalar mesons.  Note that its simple truncation already respects the Goldstone character of pions, hence MTM 
is known to work well in the ground state pseudoscalar sector. More recently, the parameters have been used more freely, and the flavour-dependent version of  
MTM has been extended to the heavy meson sector \cite{BK2011,FKW2015,HPGK2015,KGH2016,MVRB2017,KH2018}. 
In \cite{MVRB2017,GMPP2024} the spectra of heavy - light mesons were further investigated and their electromagnetic form factors were calculated. 
\cite{XU2024}.  All the above 4d BSE MTM studies of mesons were carried out in the Landau gauge, which should be particularly useful for such purposes.

 On the other hand, MTM produces a number of complex conjugated branch poles/points \cite{WI2017,DKK2015} in the quark propagator functions.   
Originally attributed to the ladder approximation, this analytical structure is instead inherent to the functional form of the MTM kernel.
Further analyses \cite{HPPW2021,HPW2023} of the gap equations in the complex plane reveal an infinite tower of branch points generated by complex conjugated poles, i.e. by the quark propagators in MTM. Unambiguously, in these models there is an infinite ladder of branch cuts traversing the complex plane transverse to the imaginary axis of the variable $q^2$.   
 In practice, this structure is ignored and sorted in a simplified fit to the MTM quark propagator. Even so, it makes the solution of BSE 
 for heavy states, and the heavier the meson considered in the MTM, the greater the uncertainties in the results.
The size of the estimated errors in masses and leptonic widths, as honestly estimated in \cite{HPGK2015,KGH2016,KH2018}, is a response to the complicated complex structure that the MTM actually has.

In an attempt to simplify the analytical properties of QCD nodes and propagators in the LRA, new models have recently been proposed.
These kernels are inspired by lattice data for which simple analytical fits have been made, and they provide the first results in an unlimited range of physical scales: 
from the pion consideration \cite{VS2020}, to the heavy masses of excited bottomonia \cite{saulibot}.

The present study takes a step back and looks at what can be achieved within the (UV improved) model \cite{saulibot} when considered in CQA. 
Instead of using the quark gap equation, the single parameter 'classical' quark propagator is used. We do not yet compare with the dressed case,
 but extend to charmonium, strangeonium and light quark systems. We discuss the presence (absence) of abnormal solution in CQA BSE. In the case of vector mesons, these states correspond to the negative norm and negative residuum BSE solution, while retaining the quantum numbers of the ordinary meson  
$C=-1, P=-1 J=1$.
The application of CQA to light mesons should be considered as a theoretical toy model. Nevertheless
the resulting spectroscopy is fascinating: We obtain the spectrum of excited mesons that agrees well with the PDG lists of experimentally known broad resonances, but their S-matrix poles are cancelled by the presence of opposite poles generated by ghost solutions. This happens regularly, suggesting that there are only cuts in the corresponding form factors in theory that should correctly describe exclusive experimental data.
 In the sections describing the results we clarify the conditions under which an unphysical uncanceled ghost could disappear from the spectrum.

\section{Quarkonia leptonic decays}

The quarkonium meson decay constant $F_V$  determines the leptonic partial decay rate $\Gamma$ of the meson with mass $M$ in the following way:

\be \label{jedna}
\Gamma_{V\rightarrow l^{+}l{-}}=\frac{\pi\alpha^2 e_q^2 F_V^2}{3M_V}(1+2r^2)\sqrt{1-4r^2} \, ,
\ee
where $\alpha$ is the fine structure constant, $e_q$ is  the quark charge  in units of the electron charge and $r$ stands for the ratio of the lepton to the vector meson
mass: $r=m_{l}/M_V$. The constant $F_V$ is defined by the matrix element of the electromagnetic current
\be
<0|j^{\mu}(0)|V(Q,\epsilon)>=F_V M \epsilon^{\mu}
\ee
where $j$ is the the electromagnetic current associated with the quark content of the meson with polarisation $\epsilon$.

 For mesons where the mixing can be ignored or idealised,  the electromagnetic  charge cab be factorised as  in the Eq.(\ref{jedna}) and
the leptonic decay can be calculated from a simple formula :
\be \label{botom}
F_{Y(n)} M_{Y(n)}=\frac{Z_2}{3} Tr_{CD}\int\frac{d^4q}{(2\pi)^4} [\gamma_{\mu} S_f(q+Q/2)\Gamma_V^{\mu}(q,Q) S_f(q-Q/2) ] 
\ee
where $S_f$ stands for given flavour quark propagator. 
The formula (\ref{botom}), as  proposed in \cite{IKR1999} and first used  in \cite{MATA1999}, 
assumes  that the meson pole part does not enter in QED Ward identity,  $\Gamma_V^{\mu}$ is thus a fully  transverse Bethe-Salpeter vertex function satisfying  
$\Gamma . Q=0$. 

The leptonic decays are hard to measure experimentaly and they have been determined for ground states and few narrow mesons \cite{KLOE2005}. No $f_V$ has been measured and  reported for excited light vectors.

\section{The BSE for quarkonium}

To obtain  vertex functions for excited quarkonia we use the BSE form as elaborated in \cite{saulibot}. 
The core of the interaction kernel  is based on lattice and DSE solutions in the Landau\&Feynman  gauge, 
but here we vary the parameters more freely  in order to obtain a
  simpler  effective description of the vector quarkonia. 
As in \cite{saulibot}, the BSE is solved  in the Feynman gauge with a regulated  gauge term.

In addition to the existing functional form of the interaction kernel  \cite{saulibot} 
we introduce a separable  effective coupling which behaves  like  $\simeq 1/ln^2 q^2 $ at large loop momenta, allowing to set the  renormalization constant $Z_2=1$ in the Eq. (\ref{botom}), as it should be for CQA. 
Since we solve the BSE without using the expansion to Tchebychev polynomials (see the discussion in \cite{MATA1999}),
 there is no need for a UV regulator   to subtract  UV divergences. They are not presented here.
For completeness we write down the  BSE kernel  here:
\be
K(k,p,Q)=\frac{4}{3}\Gamma^{\mu}(k,Q)\times \gamma^{\nu}\left[(-g_{\mu\nu}+\frac{q_{\mu}q_{\nu}}{q^2})G(q^2)
-\frac{q_{\mu}q_{\nu}}{(q^2-m_r^2)^2}\right]
\ee
with $q=k-p$ and $Q$  is the total momentum of the quark-antiquark pair. 
The transverse part is taken as following
\be \label{twopole}
G(q^2)=\frac{r_1}{q^2-m_{g1}^2}-\frac{r_2}{q^2-m_{g2}^2} \, .
\ee
These two real poles appearing in the kernel are an imperfect approximation 
 to the continuous gluonic spectral function \cite{HPRTU2021,VS2022} and it  makes the problem numerically tractable. It
  appears in the product with the following effective quark-gluon vertex
 $\Gamma^{\mu}=\gamma^{\mu}\alpha(0)\frac{\ln^2(\zeta)}{\ln^2(\zeta+k^2/\Lambda^2)}(1+\delta)$, where $\delta$ 
  is  flawour independent for ground states, but flavour dependent  otherwise. This small fluctuation  is chosen as
$\delta=\delta_0 \sin\left([\sqrt{Q^2}-M_V(1)]\frac{2\pi}{E_p}\right)$.

The numerical values for the parameters  are  taken as $\zeta=100; \Lambda=0.4 GeV ; \delta_0=0.15; E_p=1.13 GeV $
and  $M(1)$ is the ground state mass in a given spin/flavour channel, i.e. $M_{J\psi}$ and 
$M_{Y(1)}$ for the heavy quarkonia, and $M_{\phi}=1019 MeV$ $M_{\omega}=782 MeV$ for the strange and the light quark sectors. 

The infrared regulator $m_r=0.113 GeV$ has been taken identical to the first pole position $m_{g1}$, thus defining the smallest scale allowed by our consideration. This prevents the gauge term from being too large, which would otherwise produce unphysical solutions (continuous and abnormal solutions).
The position of the second pole is assumed to be $m_{g2}=E_p=1.13 GeV$ for simplicity.

 The constant mass $m_f$ appears in the constituent quark propagator: 
\be
S_f(q)=[ \not q -m_f]^{-1}
\ee
and is what constitutes the CQA for a given flavour $f$. 
In addition, the CQA ignores the momentum dependence of the renormalisation wave function $Z_f$.

The numerical value of $m_f$ is tuned to obtain a nice description of the excited states.
The ground state is treated separately to avoid unphysical states at each flavour channel. 

The CQ masses we use for this purpose are listed in Table 1.
\begin{center} 
\begin{table}
\begin{tabular}{ |c|c|c|c|}
\hline
\hline
 $ m_u $    & $ m_s $ & $m_c $& $m_b $  \\
 0.29 & 0.522     & 1.49  & 4.709    \\ 
 \hline
 \hline
\end{tabular}
\caption{\label{tabulka} Constituent quark masses (in GeV) for excited vectors. $m_u$ stands for up and down quark masses in the isospin limit.  }
\end{table}
\end{center}

We use the two-component approximation $F_1,F_5$, excluding the components responsible for the
 $D-$ wave orbital momentum. 
We already assume that the P-wave admixture due to $F_5$ is small, reducing $F_V$ by a few percent.
The leptonic decays as well as the n.c. are calculated in the $F_1+F_5$ and $F_1$ approximation. The results are shown for the latter case, assuming that $F_V$ is closer than in the case where the full set of 
is taken into account. 
   
\subsection{Abnormal states}

 When deriving the BSE and its normalisation, it is assumed that there is a real pole in the S-matrix associated with a free motion of the centre of mass of the meson. Thus, we neglect the meson decay, which  very often 
   ignores the fact that the lifetime should be much longer than the typical interaction time. This, together with other shortcomings, can lead to the occurrence of abnormal state solutions in the BSE treatment of unstable mesons.

  Obviously, the canonical Nakanishi normalisation condition 
for BSE does not fit broad resonances.
As a consequence, negative BSE 
can be seen in addition to the normal ones, and they require a negative residuum 
 as first noticed in the scalar toy model \cite{NAK1965} 
 (without relating this phenomenon to the stability of the bound states).
 
We should mention that there are more ways to violate the theory and generate unphysical states. 
To name just one: too strong renormalised coupling, if the dressing of an important Green's function is ignored (see \cite{AA1999}), etc. etc. 
However, we argue here that in QCD (and not only there),
the abnormal solution of the BSE can appear as a consequence of inappropriate
homogeneous BSE for the description of resonances. 

In the particular case of the present study, the appearance of abnormal states is somewhat more intriguing, since ghost solutions can be further supported by the CQA itself. 
 However, as we will see in the next section, the appearance of single (unpaired) ghost solutions
can be prevented by using more values for the constituent quark masses.
For $1^{--}$ neutral states, the appearance of abnormal states seems to be prevented 
by assuming only two different masses: one for the ground state and the second for the excited states.  
In fact, we found it technically easier to calculate all the excited states first
 and then refine the CQ mass to obtain a ground state without the presence of ghost solutions.
Of course, this only shows that the tuning of the constituent quark propagators is important and should not be neglected even in the heavy meson sector. 

For the purpose of the presented CQA BSE study, we numerically identify all normal and possibly all abnormal states at each flavour channel and evaluate the leptonic decay $F_V$ for both. 
We use canonical positive (and negative)
residuum normalisation condition to normalise the BSE solution for mesons (ghosts).

However, other semi-pathological or at least confusing structures can be identified.
In some cases the normal-abnormal (NA) states appear (numerically almost) degenerated. 
The poles of (NA) pairs cancel each other out.  Interestingly, when compared with experimental data 
then their appearance suggests that these pairs are an artefact of ignored coupling to decay products of resonances, 
 i.e. they are a prelude to the presence of the complex cut in more sophisticated approximations. 
To the best of the author's knowledge, the appearance of ghosts is also proper to MTM, 
however, the observation of the pairing of normal-abnormal states has either been overlooked 
or does not occur in MTM due to the very different analytical structure.
To complete the picture, there are no other known, already solved BSE solutions for vector mesons on the market.

Last but not least, it should be emphasised that all abnormal states obtained in the present study as a numerical solution can be normalised to the $-1$ residuum, i.e. they would appear as 
as negative poles in the four-point quark-antiquark functions of the form 
\be
-\frac{{\bar{\Gamma(p,P)}}\Gamma(p',P)}{P^2-M^2+\ep} \,    
\ee
thus keeping the normalisation condition for BSE ghosts identical (up to the sign) to normal states. 
We will call these negative normal states abnormal or ghosts, even though they have identical quantum numbers to normal states. 
These states are not excitations in relative time (as in the Wick-Cutkosky model \cite{NAK1965}) 
and should not be confused with them.

In order to clarify the motivations for the solutions presented in the next section, the author considers a single located abnormal state as a badness of the model approximation (e.g. inside the BSE) and we try to avoid it. On the contrary, the pairwise appearance of meson states can be regarded as a consequence of an unwise application of the homogeneous BSE, but need not be regarded as an inherent pathology due to the cancellation in the S-matrix. The position of these poles is not random, we list these cases in the next section for the sake of completeness.

\section{Results for heavy quarkonia}

The spectroscopy  and lepton decay constants as obtianed from our BSE solution are shown in the tables \ref{tabul1} \ref{tab2} for bottomonia and \ref{tab3} for charmonia respectively,
where we compare with the PDG data and several selected theoretical approaches in special case of bottomonium.

\begin{center} 
\begin{table}
\begin{tabular}{|c |c|c|c|c|c|}
\hline
\hline
name: & Y(1)      &  Y(2)  & Y(3) & Y(4) &Y(5)  \\
\hline
this work &  $9.460$      & $ 9.914\pm 10 $ &  10.36      &  $10.530 \pm 30$ & $10.740 \pm 30 $        \\
Experiment  & 9.460      & 10.23 & 10.355 &   10.579     & 10.900      \\
\hline
BSE \cite{FKW2015}  & 9.490     & 10.089  &10.327      & --- & --- \\
Salpeter \cite{WW2020} &  9.461   & 10.020    & 10.363 &  10.622     &  10.835     \\
 \hline
 \hline
\end{tabular}
\caption{\label{tabul1} Bottomonium: Comparison of calculated masses with experiment.}
\end{table}
\end{center}

For comparison with other studies, the value for $M_{Y(1)}=9.488 GeV$ was obtained in \cite{BK2011} in MTM, 
where a similar calculation was extended to excited states in \cite{FKW2015,HPGK2015}.
Recall that the parameters of the MTM for heavy quarkonia used in \cite{HPGK2015} are different from those used in \cite{FKW2015}, as are the parameters of the MTM for light mesons.  It was also shown in \cite{MVRB2017} that 
 the flavour non-universal LR interaction is required to describe the
leptonic decay constant of heavy quarkonia in the MTM.  A large uncertainty in the determination of leptonic decays
as shown in \cite{KGH2016}, is due to the numerical extraction of the bottom propagator in the MTM.

\begin{center} 
\begin{table}
\begin{tabular}{|c|c|c|c|c|c|}
\hline
\hline
name: & Y(1)      &  Y(2)  & Y(3) & Y(4) &Y(5)   \\
\hline
this work &  150     &$ 210\pm 10  $&  $ 120 \pm 5$     &  $110\pm 3$   &  $100\pm 3$     \\
\hline
Experiment  & 715      & 497 & 430 &   341     & ---      \\
\hline
BSE \cite{KGH2016} & 707     & 393  & 9(5)      & 20(15))  &----\\
\hline
SE  \cite{WW2020} &702   & 503    & 432 &  387     &---       \\
 \hline
 \hline
\end{tabular}
\caption{\label{tab2} Bottomonium: Comparison of calculated leptonic decays $F_Y$ (in MeV) with experiments}
\end{table}
\end{center}

Within the $5-7 MeV$ step, we have carefully scanned the masses above 9 GeV (and stopped at 11 GeV) and obtained all the BSE states under consideration (it is a costly procedure, the inspection of a single point takes almost 3 hours). We briefly describe our rationale in the following text:  
All narrow experimental states $Y(1)-Y(3)$ agree with some ordinary state given by the solution of our considered BSE (see table \ref{tabul1}).

In addition, we list the NA pairs.  The numerical values for the masses of the normal and abnormal states are very close, so they degenerate within the numerical errors. Three pairs of such solutions we have found are listed in the table \ref{tab22}.  
The first NA pair appears below the open botom threshold, so it should be considered  
as a slight imperfection of the BSE kernel when applied to bottomonia.

\begin{center} 
\begin{table}
\begin{tabular}{ |c|c|c|c|}
\hline
\hline
normal  &9744  & 9990       &  10346      \\
\hline
abnormal & 9793  & 10001     &  10363          \\
\hline
$f_V $(normal)  &  131     & 166 &                 \\
\hline
$f_V$(abnormal)   & 152 &   346&    \\
 \hline
 \hline
\end{tabular}e
\caption{\label{tab22}Self-excluded states. Shown are masses for   normal (first line) and abnormal states (second line) and their leptonic decay constants.  Normalization to negative unit residuum is used in order to get $f_V$ for 
abnormal states. }
\end{table}
\end{center}

The Salpeter equation is the well-known approximation for BSE,
which avoids various complications found in the more complex BSE. 
While boosting to moving frames is complicated (if possible at all), 3D reduction is particularly useful for evaluating static properties for the bottomonium system, and we provide the comparison with the study \cite{WW2020} in the tables (10-20 \% errors are not shown).

\begin{center} 
\begin{table}
\begin{tabular}{ |c|c|c|c|c|c|}
\hline
\hline
name: &$ J/\psi  $    & $ \psi(2s) $ & $\psi(3s)$ & $ \psi(4s) $ & $\psi(5s)$ \\
\hline
M (calc) &  $3097$ & $3650 \pm 10 $  & $4100 \pm 15$  & $ 4210\pm 20 $       & $ 4350\pm 30   $   \\
\hline
M (Exp)  & 3097        & 3686             &       4039 &         4153     & 4421      \\
\hline
\hline
$f_V$ (calc)  & 280          &  $120 \pm 2$ & $ 250$ & $290 \pm 20$ & $226 \pm 10$ \\  
\hline
$f_V$ (exp)  & 416          & 308              &         187   &   146            &--         \\
\hline
Salpeter \cite{WW2020} & 573 &  424 & 351  &--  & 321      \\
\hline
\hline
\end{tabular}
\caption{\label{tab3} Charmonia: Comparison of calculated masses (in MeV) and leptonic decays with experiment. }
\end{table}
\end{center}

The masses for excited $1--$ S-states charmonia  as calculated and compared to PDG data at the table \ref{tab3}.
  We do not compare with others, the reader can compare with     \cite{WW2020} for Salpeter equation, \cite{LSPB2017} for Gross equation  as well as with number of  BSE studies  \cite{FKW2015,HPGK2015,KGH2016,MVRB2017}.

Several comments are in order at this point. There is a single normal state at $3650 MeV$ corresponding to the first excited state $\psi(2s) $
to which the model was originally tuned. Then there is an AN pair at $4100$. Since these two states have different $f_V$, we keep its normal state 
in the table.  All other states are normal, there is no other abnormal state above 4 GeV, but all states appear twice, i.e. there are two close poles - two very close solutions of BSE.  The mass difference is much smaller than the width of the experimentally observed counterparts, so the pairwise appearance should be regarded as the single solution and the difference as the systematic error associated with our approach.  These errors go hand in hand with the error $\sigma^2 \simeq 10^{-7}-10^{-4}$ reached in our iteration procedure (reducing these errors would be painfully costly, note that the same level of degeneracy is seen for AN pairs). For the sake of completeness, we list the more precise positions of all the doublings observed up to 5 GeV in the Appendix. The ground state is evaluated for different values of the quark mass given below.  

\section{Light vectors and their excitations- further rise and fall of abnormal states}

The consideration and calculation methods can be extended to the light vector mesons, 
e.g. $\phi(1020)$ and its excitations modelled by the strangeonium and the $\omega/\rho$ system. 
for which we assume ideal mixing, i.e. $\omega=\frac{1}{\sqrt{2}}[uu+dd]$.
For both systems we compare with collected PDG data as presented in the decade 2010 to 2020.
 Since the LRA does not distinguish between the isospin partners $\rho$ and $\omega$, and since $\omega$ is the narrow state,
 we use the mass of omega in the phase factor in the $g$ function.  

Remember that the most accurate information about light vector resonances comes from meson production in $e^+e^-$. 
into a few mesons. Associated measured cross sections always  
include the peaks of $\omega$ and $\phi$ ($\rho$) for the odd (even) $g-$parity channel, while several 
 overlapping broad resonances are used for the phenomenological description (and are listed in the PDG) for moderately higher energies.

The BSE solutions for strangeonium are given in tables \ref{tabstran} and \ref{tab4}.
The main feature is that only the ground state is uniquely described, while all other states come with their abnormal brothers.
We list an AN pair only for the strangeonium and not for the $\omega/\rho$ system.
There are two other pairs which appear at  $1260 MeV$ and $1320 MeV$ energy.
They vanish by construction if a smaller value of mass is allowed, as will be done for the ground state.
We do not write down all the step functions that would prevent ghosts, noting that the CQA is not  adequate for the light mesons  
at all. 
  
 The optimistic feature of the BSE is that it provides (with the aforementioned constraint) a one-to-one correspondence to the 
 meson mass values listed in the PDG.  Perhaps, the unattractive feature is that the pole of the normal 
state solution is cancelled.  If they cancel out completely (assuming finite limit), then there is nothing left that is 
controversial with the appearance of ghosts in the BSE solutions for vector mesons.
  
\begin{center} 
\begin{table}
\begin{tabular}{ |c|c|c|c|c|c|c|}
\hline
\hline
Exp.  mass & $\phi(1020)$   & $X(1575)$  & $\phi(1680)$ & $\phi(1750)$ & $\phi(2170)$ &   \\
\hline
this work & 1019  &   1588   & 1960   & 2190  & 2327 & 2481   \\
\hline
 $f_V(n)$  & 230         & 80       & 119    & 147   & 166  & 190    \\
\hline
abnormal  &  --  & 1582 & 2056 & 2172 & 2302 & 2449  \\
 $f_V(a)$    &  --    &  84 &  135  &  151 & 173 & 204 \\
 \hline
 \hline
\end{tabular}
\caption{\label{tabstran} Strangeonium masses compared against PDG list of  excited $\phi$ (and one X ) mesons.
 Negative norm/residuum states and their decay constants are displayed  in 4-th and 5-th line respectively. 
Ground state is solved separately as explained in the text.} 
\end{table}
\end{center}

\begin{center} 
\begin{table}
\begin{tabular}{ |c|c|c|c|c|c|}
\hline
\hline
Exp.  mass & $\rho(770)$   & $\rho(1450)$  & $\rho(1570)$ & $\rho(1700)$ & $\rho(2150)$    \\
\hline
Exp.  m. (singlet) & $\omega(782 )$ & $\omega(1420)$   & $\omega(1650)$  & $ X(1750)$  & $\omega(2205)$     \\
\hline
this work  & 782     & 1400         & 1564 \& 1650        & 1750 \& 1881         &   2015      \\
\hline
$f_V $ & 36 & $109 \pm 20$ & $107\pm 30$ & $135\pm 20$  & $ 160 \pm 30 $         \\  
 \hline
 \hline
\end{tabular}
\caption{\label{tab4}  Masses of $q\bar{q}$ (third line) compared to masses of light $1^{--}$ resonances as listed in PDG lists 2012-2020
 (the first and the second line). Calculated decay constants values are shown  in the forth line. 
Negative norm states are not displayed and the ground state is calculated for different constituent propagator  
as explained in the text.}
\end{table}
\end{center}

\subsection{Getting rid of abnormal states}

  Except in one case, not a single ghost appears among the excited states. So to prevent them completely means 
is to find a different CQ mass value or perhaps a different constant $Z_f$ that prevents ghosts near the ground state.
Varying only the mass, we have found that this is possible for bottomonia and for the $\phi$ meson, although 
leptonic decay is still not ideal. We do not fine-tune the model to get the best agreement with the experiment, but wait for changes and improvements,
after DSE and BSE are solved simultaneously.
The leptonic widths calculated in all the tables are the results of our simple search.   
To get the $\omega$ in the given model and $J\Psi$ without a ghost nearby, one must also take into account $Z_f$ or alternatively further increase the infrared coupling strength. We expect this to be achieved naturally in the simultaneous DSE-BSE treatment.  

 Even in the case where changing one constant is enough to prevent ghosts, the solution is not without ambiguity.
Keeping the mass for the ground state fixed,
we have found at least two possible values for the CQ mass of bottomonium $9460$, the first value is $m_b^{Y1}=4.42 GeV$ and the second, for which we also get the bound state mass, is $m_b^{Y1}=3.37 GeV$.
The smaller mass solution gives a decay of $f_V=150 MeV$,
which is still a long way from the experimental value, but is a better fit.

The same does not happen with charmonium. While we get rid of the ghost solution that we would otherwise face at $3150 MeV$, 
 then it is replaced by the NA pair, both components weighting $3097 MeV$ for a constant quark mass $m_c^{J/\Psi}=1.263 GeV$ for the 
normal state and $m_c^{J/\Psi}=1.266 GeV$ for the ghost.  

In the case of the strangeonium, scanning the CQ mass for the $\phi$ meson while keeping the BSE interaction intact, we also get more eigenvalues. 
The first is the ghost that appears for the CQ mass $395 MeV$, then the pair of normal solutions for
$ m_s=342MeV,343.5 MeV$, giving almost identical leptonic decay constants $f_V=232MeV$ and $f_V=234 MeV$ respectively. The renormalisation function was $Z_f=1$ (i.e. no needed to consider to get the results).  

To conclude this section, we find the solution for the physical mass of $\omega$,
While no single (unpaired) solution normal state was found for excited states, but single physical  solution
can be identified if the constituent qurk mass is bellow  $290 MeV$. There are more values of CQ masses that fit the $\omega$ mass.
 In fact, the most numerically precise solution is obtained with $m_u=108 MeV$. Obviously this CQ mass value is rather smaller than the usually assumed constituent quark mass. We also get unreliable deacy constant  $F_{\omega}=36 MeV$ and $F_{\omega}=34 MeV$ for single and two components approximation respectively.  Note also, that in this exceptional case we have used $Z_f=1/1.25$ to obtain presented nontrivial  result. We report this unreliable and in fact worng  values as a curiosity, just to inform the reader that  CQM BSE solutions exist in this channel. Let us stress, that to get more uniform picture consistent with other spin channels (e.g with pion) one  needs to go beyond CQM and incorporate both ignored ingrediences: the running mass and momentum dependent renormalization function.
We hope we report such  result in the near future.

\section{Conclusion}  
      
We have calculated the mass spectrum and the leptonic decay constant of flavourless vector mesons using CQ LRA BSE in the Feynman gauge.
 There is a striking difference between the description of the heavy quarkonia and the excited mesons of the light (anti)quarks.  In the latter case, the solution for the normal excited state regularly comes in pairs with the abnormal state. 
Getting a good (normal) solution for the narrow ground state and facing the cancellation between normal and abnormal states 
 in these pairs, we argue that this is due to the inconsistent use of the BSE to describe broad resonances.

On the other hand, if one admits that the NA pairs are an artefact of unwise use rather than bad approximations within the BSE, 
one obtains a remarkable description of excited heavy quarkonia. 

Another shortcoming of the single mass value CQ BSE is either the bad agreement with the experiment and the appearance of the uncanceled ghost 
in the vicinity of the ground state. These ghosts have no good meaning,
and they can be eliminated by using slightly different
 CQ mass in the propagators of the BSE.In the case of heavy quarkonia, 
the values of the CQ masses used for the excited and ground states 
 are proportional to the ratio

\be
\frac{m_c^{ground}}{m_c^{excited}}=\frac{V(1)}{V(2)}
\ee
 
 which is nothing more than a prelude to the need for dynamical mass. 

In fact, we expect that there will be no single ghost state in a more consistent DSE/BSE treatment.
On the other hand, we expect that an AN pair can in principle appear even in more advanced approximations. 

On the other hand, the use of a single interaction kernel in the BSE description of variety mesons is remarkable. 
We expect that after taking into account the dressing of the propagators, we will also get a good description in other spin channels.
These and similar advanced approaches \cite{SA2022,MIBA2023} describe resonances in a given exclusive process. They offer alternatives to popular (but more complicated) coupled channel analyses (see for example \cite{HMS2022,LIO2020}) without the use of effective models but within the gluon and quark degrees of freedom.

\appendix
\section{Normalization of BSE}

Performing trivial integrations, making a trace,  the normalization used  for purpose of  our numerics  reads:

\bea
\pm 1&=&\frac{N_c}{2\pi^3}\int_{-\infty}^{\infty} d q_4 \int_{0}^{\infty} d q_s q_s^2 I
\nn \\
I&=&\frac{q_4^2-\Delta}{[(q_4^2+\Delta)^2+q_4^2 M_V^2]^2}[F_1^2(3m_f^2+\frac{3}{4}M_V^2+q_E^2+2q_4^2)
\nn \\
&+&q_s^2 F_5^2(q_E^2+\frac{M_V^2}{4}-m_f^2)-4 m q_s^2 F_1 F_5]
\nn \\
&-&\frac{1}{(q_4^2+\Delta)^2+q_4^2 M_V^2}(\frac{3}{2}F_1^2+\frac{1}{2} F_5^2 q_s^2)+...
\eea
where ellipses stands for less then $1\%$ contribution due to the derivative of the interacting kernel and where
$q_E^2=q_4^2+q_s^2$ and
\be 
\Delta=q_s^2+m_f^2-\frac{M_V^2}{4} \, .
\ee
where plus and minus stands for the normal and abnormal state respectively.
In shallow binding limit $2m_c \rightarrow M$ the denominator is small but the $q_4^2-\Delta$ can abruptly change the sign.  
The result depends on a way how it is weighted by $F_1^2$.  An interplay decides between normal and abnormal state.
  
A large number of integration points $288*144$  were used to integrate 2d BSE. 
A large number of mesh points is also needed to eliminate false solutions that would otherwise appear.  

\section{Detailed list of numerical results   for charmonia}

Observed  solution pairs for BSE for charmonia are listed in the table 
\ref{tabpair}. Only the first one is an NA pair, all others are ordinary states. Assuming exact degeneracy, the systematic error is determined from the mass  difference 
for each pair, so  not worse than  1\%. 
Leptonic widths are more sensitive, mainly to the integral error represented by the norm. In a few cases we have used $20 GeV$ UV cut-off
to avoid from numerical noise at the UV tails of the vertex functions. 
\begin{center} 
\begin{table}
\begin{tabular}{ |c|c|c|c|c|c|c|c|}
\hline
\hline
M  &  $4084^g ; 4100 $ & 4200; 4226   &4341; 4368  & 4495; 4527   &  4670; 4702  & 4865; 4899 & 5089; 5119    \\
\hline
F  & $350^g$ ; 250 & 309;270                  & 230; 223    & 186; 181      & 162; 160      & 140; 144     &  132;130 \\
\hline
\hline
\end{tabular}
\caption{\label{tabpair} Eigenvalues  of the  BSE and leptonic decays as obtained from the BSE  for vector 1S charmonia in details.
 $g$ stands for negative norm state.}
\end{table}
\end{center}

\end{document}